\documentclass[twocolumn]{article}

\newcommand{\bbbone}{{\mathchoice {\rm 1\mskip-4mu l} {\rm 1\mskip-4mu l}
{\rm 1\mskip-4.5mu l} {\rm 1\mskip-5mu l}}}

\date{}
\title{Teleportation of entangled states and dense coding
using a multiparticle quantum channel}

\author{V.N. Gorbachev, A.I. Trubilko, A.I. Zhiliba, E.S. Yakovleva}
\begin{document}
\maketitle

\begin{abstract}
A set of protocols for teleportation and dense coding tasks with the use of a $N$
particle quantum channel, presented by entangled states of the GHZ class, is
introduced, when $N>2$. Using a found representation for the multiparticle entangled
states of the GHZ class, it has shown, that for dense coding schemes enhancement of
the classical capacity of the channel due from entanglement is $N/N-1$. If $N>2$ there
is no one-to-one correspondence between teleportation and dense coding schemes in
comparison with the EPR channel is exploited. A set of schemes, for which two
additional operations as entanglement and disentanglement are permitted, is considered.
\end{abstract}

\section{Introduction}

The large number of the quantum information tasks are based on a channel represented
by entangled states. An EPR pair or a two particle quantum channel is a main resource
for dense coding and teleportation to be attractive now for many applications. Dense
coding has been introduced by Bennett et al \cite{BenWdCod} and demonstrated in the
optical experiment with polarized photons by Zeilinger et al \cite{ZeidCod} and for
continuous variables by Peng et al \cite{PengdCod}. The quantum teleportation protocol
proposed by Bennett et al \cite{TelBen} has been implemented by several groups
\cite{TelExp}.

A lot of teleportation and dense coding schemes, including its applications has been
considered by many authors both for discrete and continuous variables. Recently an
attempt to classify all schemes has been made by Werner \cite{WeR}. Generally, this
problem is very difficult and the main results have been obtained in the case of
so-called \emph{tight} schemes realized with minimal resources. It has been found a
one-to-one correspondence between all \emph{tight} schemes of teleportation and dense
coding. It is worth noting that an EPR pair as a two-particle quantum channel is used
in the \emph{tight} schemes. The obtained result is significant for practice because
it tells, that  if one can teleport a qubit, then he can perform dense coding using
the same experimental arrangement without any additional resources.

In this paper a quantum channel presented  by a multiparticle  entangled state of the
GHZ class is considered for teleportation and dense coding. When the channel involves
more then two particles, its features becomes  quite complicated and all schemes are
not \emph{tight}. For particular case, one finds the GHZ channel based on the triplet
of the GHZ form. If we wish to exploit the GHZ channel for teleportation, for example,
then the task can't be accomplished by a generalization of the usual protocol simply.
Considering how to transmit an unknown qubit by the GHZ channel, Karlsson et al have
shown, that the unknown state can be recovered by one of the two receivers, but not
both \cite{Kar}. What kind of a two-qubit state can be teleported through the GHZ
channel? This problem has been considered by Marinatto et al \cite{Luca M}. It has
been found that the general two-qubit state can't be transmitted perfectly but a pure
entangled states can. This conclusion is in agreement with result obtained in Ref.
\cite{VA}. With the use of the GHZ channel, the conditional teleportation of two
entangled qubits has developed \cite{China}.

The main goal of this work is to consider the multiparticle quantum channel for
informational tasks as  teleportation and dense coding. In a particular case of the GHZ
channel a scheme for distributing a mixed qubit state with two parties is discussed.
The set of the questions,  we study in this paper, is the following: what is the dense
coding schemes, whether we could have an enhancement of the channel capacity, whether
the teleportation resources can be used directly for dense coding similarly
\emph{tight} schemes, what kind of teleportation and dense coding schemes can be
created using certain additional resources such as entanglement and disentanglement
operations.

The paper is organized as follows.  First, we discuss the main resources and consider
\emph{tight} schemes, then teleportation and dense coding protocols are introduced for
the GHZ channel and a telecloning scheme is presented. In the next section using the
found representation of $N$ particle entangled states we establish some main features
of the multiparticle channel and calculate its capacity. Then a collection of
teleportation and dense coding schemes is briefly discussed when such operations as
entanglement and disentanglement are permitted.

\section {Tight schemes}
Following to Werner \cite{WeR}, we consider a set of objects to create some
teleportation and dense coding schemes. The set includes an observable $F$, a
collection of unitary operators $T$, an entangled state $\omega$ to be a quantum
channel. Let the Hilbert spaces of the involved systems have the same dimension $d$,
and $\omega$ is the $N$ particle entangled state. Two parameters $d$ and $N$ play the
key role. If $N=2$, one can find schemes called \emph{tight}.

Let the observable $F$ be a complete set of the $N$- particle states,
$\sum_{x}F_{x}=1$, $F_{x}=|\Phi_{x}(N)\rangle \langle \Phi _{x}(N)|$, where $x$ is one
of the $d^{N}$ elements of an output parameter space $X(d^{N})$. In general these pure
states can be not maximally entangled. We assume, that $T$ is the collection of the
$m$ - particle unitary operators $U_{x}(m)$, completely positive, that transform  input
state of the channel $\omega$ to output state $U_{x}(m)\omega U^{\dag}_{x}(m)$. Let the
$N$ - particle quantum channel $\omega=|\Omega\rangle\langle\Omega|$ be shared $N$
parties $A, B, C,..$, spatially separated, where $\Omega$ can be one of the states of
$\Phi_{x}(N)$. Then all resources are
\begin{equation}\label{11}
R=\{\omega,  x\in X(d^{N}), \Phi_{x}(N), U_{x}(m)\}
\end{equation}
Using (\ref{11}) the teleportation and dense coding schemes can be obtained. We
consider only the qubit case, for which $d=2$. Note, the multiparticle quantum channel
has new properties due from operators $U_{x}(m)$.  When $m\geq 2$, these operators may
be non local and the main resources (\ref{11}) is not a set of the LOCC in contrast the
\emph{tight} schemes.

If $N=2$, one finds a two-particle quantum channel, represented by the EPR state, say
of the form $ \Omega=(|00\rangle+|11\rangle)/\sqrt{2}$. The channel is shared two
parties, Alice and Bob. Here the observable $F$ is described by the Bell states
$\Phi_{x}(2)=\Phi^{+},\Psi^{+}, \Phi^{-},\Psi^{-}$ and the set of unitary operators
consists of the Pauli and the identity operators $U_{x}(1) =\bbbone, \sigma_{z},
\sigma_{x}, -i\sigma_{y}$, where $\sigma_{y}=i\sigma_{x}\sigma_{z}$. In this case the
space $X$ has four elements $x=0,1,2,3$ by which the 2 bits of information can be
encoded. The following map is possible
\begin{equation}\label{12}
x\leftrightarrow \Phi_{x}(2)\leftrightarrow U_{x}(1)
\end{equation}
or in more details
\begin{equation}\label{13}
\begin{array}{cccc}
x&\tilde{x}   &\Phi_{x}(2) &U_{x}(1)\\
0&00          &\Phi^{+} &\bbbone\\
1&01          &\Psi^{+} &\sigma_{z}          \\
2&10          &\Phi^{-} &\sigma_{x}\\
3&11          &\Psi^{-} &-i\sigma_{y}
\end{array}
\end{equation}
where $\tilde{x}$ is encoded by binary notation of $x$.

Reading (\ref{12}) as $\Phi_{x}(2)\mapsto x\mapsto U_{x}(1)$ one finds teleportation,
that allows Alice sending to Bob  an unknown qubit
\begin{equation}\label{14} \zeta=\alpha|0\rangle+\beta|1\rangle
\end{equation}
where  $|\alpha|^{2}+|\beta|^{2}=1$. In accordance with the teleportation protocols,
Alice performs the Bell state measurement on her half of ERP pair and the unknown
qubit. Outcomes of the measurement  $x$ can be encoded  with the use of the unitary
operators $U_{x}(1)$, by which Bob acts on his half of EPR pair to recover the unknown
state. One ERP pair and the 2 bits of classical information are needed for teleporting
a single qubit.

Reading (\ref{14}) as $x\mapsto U_{x}(1)\mapsto \Phi_{x}(2)\mapsto x $, one can find
the dense coding scheme, that permits  Alice sending of a two-bit message to Bob,
manipulating one bit only.  Because of the 2 bits of information
$\tilde{x}=00,01,10,11$ can be encoded by the four operators $U_{x}(1)$, Alice can
generate the Bell basis acting on her particle of EPR pair
\begin{equation}\label{15}
\Phi_{x}(2)=(\bbbone\otimes U_{x}(1))\Omega
\end{equation}
Then the 2 bits of information are storied in four orthogonal states, that can be
distinguished, if Bob performs the Bell state measurement. The properties of this
channel can be described by the Holevo bound, that tells us that the classical
capacity of this quantum channel increases twice because of entanglement.

The considered schemes of teleportation, which is perfect, and dense coding are called
\emph{tight} \cite{WeR} in the sense of the required resources. These resources are the
EPR channel, the 2 bits of information, the Bell state measurement and a collection of
the one-particle unitary operators. Werner has shown that for all tight schemes there
is a one-to-one correspondence between teleportation and dense coding.

\section{The GHZ channel}

If three particle entanglement is used instead of  EPR pair one finds a channel which
features are more complicated. This channel shared multi users, Alice, Bob and Claire,
allows not only transmitting of quantum and classical information by teleportation and
dense coding, but  distributing quantum states between several parties by coping or
telecloning.

There is a complete set of the three particle entangled states of the form
\begin{eqnarray}\label{21}
(|000\rangle \pm |111\rangle)/\sqrt{2},~~ (|001\rangle \pm |110\rangle)/\sqrt{2}&&
\nonumber
\\
(|010\rangle \pm |101\rangle)/\sqrt{2},~~  (|011\rangle \pm |100\rangle)/\sqrt{2}&&
\end{eqnarray}
Without loss of generality a triplet of the GHZ form  can be chosen as the quantum
channel, whose three particles $A,B$ and $C$ are shared Alice, Bob and Claire. Then
$\Omega=|GHZ\rangle$, where
\begin{equation}\label{22}
|GHZ\rangle=\frac{1}{\sqrt{2}}(|000\rangle+|111\rangle)_{ABC}
\end{equation}
All schemes  based on the GHZ channel, are not tight and the theorem by Werner can't
guarantee a one-to-one correspondence between teleportation and dense coding schemes.

\subsection{Teleportation}

One of the main features of the GHZ channel is perfect transmitting of the two
particle entangled states of the EPR form
\begin{equation}\label{23}
\zeta=\alpha|01\rangle+\beta|10\rangle
\end{equation}
where $|\alpha|^{2}+|\beta|^{2}=1$. It has been shown by Marinatto and Weber, that the
general state of a pair of qubits cannot be transmitted through the GHZ
channel\cite{Luca M}.

If Alice wishes to send the entangled state $\zeta$ of the qubit 1 and 2, she needs to
perform the measurement on  particles 1, 2 and her particle $A$ of the GHZ channel. The
task can be accomplished, if the observable $\Phi_{x}(3)$  has the form of product of
a superposition state of particle 1, say $\pi^{\pm}=(|0\rangle\pm
|1\rangle)/\sqrt{2}$, and the Bell states of particles 2 and A. This basis is
described by states in which two particles only are maximally entangled,  it has the
form
\begin{equation}\label{24}
\Phi_{x}(3)=\{\pi^{\pm}_{1}\otimes \Phi^{\pm}_{2A},  \pi^{\pm}_{1}\otimes
\Psi^{\pm}_{2A}\}
\end{equation}
The measurement given by  (\ref{24})  projects  particles $B$ and $C$ into the state
that is connected by unitary transformation with the unknown state
\begin{equation}\label{25}
|BC\rangle_{x} =U_{x}(2)|\zeta\rangle
\end{equation}
where $|BC\rangle_{x}=\langle \Phi_{x}|\zeta\rangle \otimes |GHZ\rangle/Prob(x)$, and
probability of all outcomes $Prob(x)=1/8$.

The unitary operators can be chosen in the factorized form
\begin{equation}\label{26}
U_{x}(2)=B_{x}\otimes C_{x}
\end{equation}
where $B_{x}$ and $C_{x}$ acts on the  Bob and Claire particle. It means that Alice
cannot rotate the qubit of Bob and vice versa, but transformations are correlated
because of they have the same indexes. It is a case of LOCC. In more details  the
presented protocol reads
\begin{equation}\label{27}
\begin{array}{rcrcc}
x&\Phi_{x}(3)               &|BC\rangle_{x}                           &B_{x}       &C_{x}\\
0&\pi^{+}\otimes\Phi^{+} & \beta|00\rangle +\alpha|11\rangle &  \sigma_{x}& 1      \\
1&\pi^{+}\otimes\Phi^{-} & \beta|00\rangle -\alpha|11\rangle & i\sigma_{y}& 1      \\
2&\pi^{-}\otimes\Phi^{+} &-\beta|00\rangle +\alpha|11\rangle &-i\sigma_{y}& 1      \\
3&\pi^{-}\otimes\Phi^{-} &-\beta|00\rangle -\alpha|11\rangle & -\sigma_{x}& 1      \\
4&\pi^{+}\otimes\Psi^{+} & \beta|11\rangle +\alpha|00\rangle & 1          &  \sigma_{x}\\
5&\pi^{+}\otimes\Psi^{-} & \beta|11\rangle -\alpha|00\rangle & 1          & -i\sigma_{y}\\
6&\pi^{-}\otimes\Psi^{+} &-\beta|11\rangle +\alpha|00\rangle & 1          &  i\sigma_{y}\\
7&\pi^{-}\otimes\Psi^{-} &-\beta|11\rangle -\alpha|00\rangle &
1          &-\sigma_{x}
\end{array}
\end{equation}
Two points may be made about it. First, in (\ref{27}) there are only four vectors
$|BC\rangle_{x}$ which describe four different physical states of a system. So that,
outcomes $x=0$ and $x=3$, result in the states $\beta|00\rangle +\alpha|11\rangle$ and
$-(\beta|00\rangle +\alpha|11\rangle)$ to be equal up to a phase factor, that has no
physics reason. Indeed, they can be obtained by different way, for example
$\beta|00\rangle_{BC}
+\alpha|11\rangle_{BC}=(\sigma_{x}\otimes\bbbone)|\zeta\rangle_{BC}=
(i\sigma_{y}\otimes\sigma_{z})|\zeta\rangle_{BC}$. The second is more important. The
simple observation shows that the task can be accomplished, if operators $U_{x}(2)$
are not factorized. For the considered outcomes, say $x=0$, one finds
$(\sigma_{x}\otimes\bbbone)C_{BC}C_{CB}C_{BC}|\zeta\rangle_{BC}=\beta|00\rangle_{BC}+\alpha|11\rangle_{BC}$,
where $C_{ct}$ is CNOT operation, $c$ is a control bit, $t$ is a target bit, $c,t=B,C$.
This case results in the non-local operations.

Teleportation of the considered two particle entangled state can be achieved by the
usual protocol, that recommends to do it one-by-one. It needs two EPR pairs instead of
the GHZ channel, which  seems to be less expensive in comparison with two entangled
pairs.

\subsection{Dense coding}
Is it possible to use the above teleportation resources given by (\ref{27}) for dense
coding similar the case of the EPR channel? The answer is not, but a scheme of dense
coding can be achieved.

Let a sender wishes to transmit a three bit message.
The 3 bits of information $000,
001, \dots, 111$ can be encoded by a set of the eight states $D_{x}$ each of which is
obtained from the GHZ state using a collection of the unitary operators $U_{x}(2)$ in
accordance with (\ref{15}), for example. Let the two particle operators be chosen
factorized in the form (\ref{26}), then equation (\ref{15}) reads
\begin{equation}\label{28}
|D_{x}\rangle_{ABC}=\bbbone\otimes B_{x}\otimes C_{x}|GHZ\rangle_{ABC}
\end{equation}
An appropriate collection of the Pauli operators permits the sender to generate the
complete set of states $D_{x}$, given by (\ref{21}). All these states are well
distinguishable by measurement, which outcomes encode the three bit message. Then one
finds a dense coding scheme, that is described by the map of the form
\begin{equation}\label{29}
\begin{array}{ccccc}
\tilde{x}  & B_{x}       &C_{x}        & D_{x}=B_{x}C_{x}|GHZ\rangle_{ABC}    \\ 
000&  \bbbone     &\bbbone      & |000\rangle+|111\rangle  \\
001&  \bbbone     &\sigma_{x}   & |001\rangle+|110\rangle  \\
010&  \sigma_{z}  &\bbbone      & |000\rangle-|111\rangle  \\
011&  \sigma_{z}  &\sigma_{x}   & |001\rangle-|110\rangle  \\
100&  \sigma_{x}  &\bbbone      & |010\rangle+|101\rangle  \\
101&  \sigma_{x}  &\sigma_{x}   & |011\rangle+|100\rangle  \\
110& -i\sigma_{y} &\bbbone      & |010\rangle-|101\rangle  \\
111& -i\sigma_{y} &\sigma_{x}   & |011\rangle-|100\rangle
\end{array}
\end{equation}
In (\ref{29}) we have omitted  the normalization factor $1/\sqrt{2}$ in $D_{x}$.

To find the capacity of this GHZ channel it needs to calculate the Holevo function
\begin{equation}\label{29H}
C(\{p_{i}\},\rho)=S(\rho)-\sum_{i}p_{i}S(\rho_{i})
\end{equation}
where $\rho=\sum_{i}p_{i}\rho_{i}$, $\rho_{i}$ are the density matrices of the states
sent to receiver according to probabilities $p_{i}$ and $S(\rho)$ is the von Neumann
entropy. For the considered case $\rho_{i}=|D_{i}\rangle\langle D_{i}|$ and the
channel is represented by the maximally entangled state, then assuming $p_{i}=1/8$,
one finds $C=S(\sum_{i}|D_{i}\rangle\langle D_{i}|/8)=3$, hence
 per transmitted bit $C/2=3/2$,
that is the classical capacity of the quantum channel. It means, the channel capacity
due from entanglement increases in 3/2 times. However, this result is clear without
any calculations. Because of the presented protocol allows sending the 3 bits of
information manipulating two bits only, then profit is 3/2, which is enhancement of
the capacity. Also it is clear, that it results from the entanglement, which degree
has to be maximum.

Inspection of (\ref{27}) shows that the teleportation resources are inapplicable for
dense coding. The reason is that the states $D_{x}$ obtained in accordance with
equation (\ref{28}), where operators $C_{x}, B_{x}$ are given by (\ref{27}), is not a
\quad complete set. Also, being suitable for dense coding the complete set given by
(\ref{21}) can't be used for teleportation because of outcomes of measurement depend
on the unknown state. Therefore there is not a one-to one correspondence between these
schemes as for \emph{tight} ones. However a connection can be established. Indeed, two
sets $D_{x}$, given by (\ref{26}), and $\Phi_{x}(3)$ denoted by (\ref{27}) can be
transformed from one to another by the unitary operation, say of the form
$\Phi_{x}=H_{B}C_{BC}D_{x}$, where $H_{B}$ is the Hadamard transformation of particle
$B$. It follows from (\ref{28}), that eight distinguishable states can be obtained by
manipulating only two bits of the GHZ channel as follows
\begin{equation}\label{281}
\Phi_{x}(3)=H_{B}C_{BC}(\bbbone\otimes B_{x}\otimes C_{x})|GHZ\rangle
\end{equation}
where $B_{x}$, $C_{x}$ are given by (\ref{29}).

The equation (\ref{281}) tells, that the measurement from the teleportation scheme may
be used for dense coding. For that it needs to replace the operations $B_{x}\otimes
C_{x} \to H_{B}C_{BC}(B_{x}\otimes C_{x})$ before sending the message. Then the 3 bits
of information are stored in the complete set of states to be well distinguished by
the projective measurement of $\Phi_{x}$. Note, the unitary operations become non
local, what is the one of the particular qualities of the three-particle channel.

Indeed, for dense coding schemes the GHZ channel can be created by operations
$U_{x}(2)$. Let only two qubits $A$ and $B$ of the three ones $A,B$ and $C$ be
entangled, in other words the EPR channel and the ancilla qubit $C$ are introduced,
then the GHZ state can be prepared by the way
$C_{BC}(|\Phi^{+}\rangle_{AB}\otimes|0\rangle_{C})=|GHZ\rangle_{ABC}$. This
transformation can be inserted into each unitary operator $U_{x}(2)$, that becomes
more complicated because of $U_{x}(2) \to U_{x}(2)C_{BC}$. In the same time it looks
as the EPR pair is used instead of the GHZ channel.

\subsection{Telecloning}
The GHZ channel shared three parties $A$,$B$ and $C$, spatially separated, allows
distributing information with $B$ and $C$. Two copies of an unknown state can be
produced by a teleportation protocol so that we shall call it telecloning. We consider
a scheme,  whose main resources are the Bell state measurement and set of the Pauli
operators as for \emph{tight }schemes.

Let using the above resources Alice wishes to sent to Bob and Claire an unknown qubit
in a mixed state
\begin{equation}\label{30}
\rho_{1}=\lambda_{0}|0\rangle\langle 0|+\lambda_{1}|1\rangle
\langle 1|
\end{equation}
Then  combined state is $\rho_{1}\otimes|GHZ\rangle\langle GHZ|$. After the Bell state
measurement on the unknown qubit 1 and the qubit $A$ from the $GHZ$ channel the
reduced density matrix of particles $B$ and $C$ has the form
$\rho_{BC}=\lambda_{0}|bb\rangle\langle bb|\pm\lambda_{1}|\bar{bb}\rangle\langle
\bar{bb}|$, where $b=0,1$, $\bar{b}=1-b$. Two bits of information allows Bob and
Claire perform the local unitary operations to obtain the state
\begin{equation}\label{305}
\rho_{BC}'=\lambda_{0}|00\rangle\langle 00|+\lambda_{1}|11\rangle\langle 11|
\end{equation}
One find $\rho_{BC}'$ to be a separable and classically correlated state. In the same
time both receivers have in their hands the unknown state, since the reduced matrices
read  $\rho_{B}=\rho_{C}=\lambda_{0}|0\rangle\langle 0|+\lambda_{1}|1\rangle \langle
1|$.

As result, two perfect copies of an unknown mixed state can be made by teleporting.
Note, these copies are not independent, that follows from the non cloning theorem.
However each receiver can manipulate his state independently, if and only if he
performs local unitary operations. It is not true, when one of them decides to make a
measurement of his state.

\section{The N-particle quantum channel}

Some main features of the teleportation and dense coding schemes can be summarized,
considering a multi particle channel, for which the following mapping plays the key
role
\begin{equation}\label{40}
  x \leftrightarrow U_{x}(m)\leftrightarrow \Phi_{x}(N)
\end{equation}
By contrast the \emph{tight} schemes, it seems to be a hard problem to proof it
generally, therefore we will restrict several facts.

\subsection{Representation for multiparticle states of the GHZ class}

Considering resources given by (\ref{11}) one finds the factor $m$ in operator
$U_{x}(m)$ to be important, as it might be noticed from the GHZ channel. For
teleportation schemes $m$ is a number of particles on which an unknown state is
transmitted, in other words, $m$ shows how many particles can be teleported by the
channel. For dense coding  $m$ indicates a number of particles for manipulating to send
the $N$ bit message and ratio $N/m$ becomes the classical capacity of the quantum
channel due from entanglement.

A one-to-one correspondence $x \leftrightarrow \Phi_{x}(N)$, where $x$ is one of the
$2^{N}$ outcomes of the von Neumann measurement, described by a complete but not over
complete set of projectors, is clear. By contrast the map $U_{x}(m)\leftrightarrow
\Phi_{x}(N)$ is not so trivial. Similarly (\ref{15}) one can write
\begin{equation}\label{41}
\Phi_{x}(N)=( \bbbone^{\otimes (N-m)}\otimes U_{x}(m))\Omega
\end{equation}
where $\bbbone^{\otimes (N-m)}$ is tensor product of $N-m$ identity operators
$\bbbone\otimes \bbbone...$. According to the following rough dimension count, factor
$m$ can be established from (\ref{41}). In fact, being the $N$ qubit state, vector
$\Phi_{x}(N)$ has the $2^{N}$ components. Any  the $m$ qubit operator $U_{x}(m)$ has
the $2^{2m}$ matrix elements. Then for correspondence between $\Phi_{x}(N)$ and
$U_{x}(m)$, it needs
\begin{equation}\label{42}
m\geq \frac{N}{2}
\end{equation}
Indeed, these reasons are true not only in the qubit case, but for arbitrary dimension
of the Hilbert space $d$.

A simple observation allows to obtain the factor $m$ with more accuracy.
 Let the channel be represented by a maximally entangled state $\Omega$
of the GHZ class
\begin{equation}\label{43}
|\Omega\rangle =\frac{1}{\sqrt{2}}(|0\rangle^{\otimes N}
+|1\rangle ^{\otimes N})
\end{equation}
where $|b\rangle ^{\otimes N}$ is tensor product $|b\rangle\otimes\dots|b\rangle$,
that is a state of the $N$ independent qubits in the Hilbert space
$\mathcal{H}_{1}\otimes\dots\mathcal{H}_{N}$, $b=0,1$.

\textbf{Preposition.} \emph{The set of the N particle vectors
\begin{eqnarray}\label{44}
\Phi_{b_{1}b_{2}\dots b_{N}}(N)=&&\\
\nonumber \frac{1}{\sqrt{2}} (|0\rangle\otimes|b_{2}\dots
b_{N}\rangle+(-1)^{b_{1}}|1\rangle\otimes | \bar{b}_{2} \dots\bar{b}_{N}\rangle)&&
\end{eqnarray}
from the Hilbert space $\mathcal{H}_{1}\otimes\dots\mathcal{H}_{N}$, where $b_{1}=0,1$
$|b_{2}\dots b_{N}\rangle=|b_{2}\rangle\otimes\dots|b_{N}\rangle$,
 and $|b_{k}\rangle =0,1$,  $\bar{b}_{k}=1-b_{k}$  is the orthonormal
basis in $\mathcal{H}_{k}$ for each $k=2,\dots N$, is the complete set of maximally
entangled states}.

If $N=2$, one finds the Bell states $\Phi_{b_{1}b_{2}}(2)=(|0~
b_{2}\rangle+(-1)^{b_{1}}|1~ \bar{b}_{2}\rangle)$, that are generated by two classical
bits $b_{1}, b_{2}=0,1$. When $N=3$ three bits $b_{k}=0,1$, $k=1,2,3$ generate the set
$D_{x}(3)=\Phi_{b_{1}b_{2}b_{3}}(3)$, given by (\ref{21}). Also $\Omega
=\Phi_{00\dots0}(N)$ belongs to the collection  (\ref{44}).

{\bf{Proof}}. Each of the states, that has the form (\ref{44}), is maximally entangled
in the sense of the reduced von Neumann entropy $E=S(\rho(1))$, where $\rho(1)$ is the
one particle density matrix. It follows from (\ref{44}), that for any particle
$\rho(1)=\bbbone /2$, then $E=1$, and entanglement is maximum. Also one finds the
considered set of states to be complete, because of condition
\begin{equation}\label{aa}
\sum_{b_{1}\dots b_{N}=0,1}|\Phi_{b_{1}b_{2}\dots b_{N}}(N)\rangle\langle
\Phi_{b_{1}b_{2}\dots b_{N}}(N)|=1
\end{equation}
that directly results from the completeness of the collections  $|b_{k}\rangle$.

Note, all possible entangled states can't be written in the form (\ref{44}). It
represents the GHZ like class only and, for example, W - states introduced by Cirac et
al \cite{W} and ZSA (Zero Sum Amplitude) - states proposed by Pati \cite{ZSA}, that
can't be transformed from the GHZ states by local operations, have another form.

The next observation plays the key role.  Equation (\ref{44}) tells, that to generate
all states of the set, it needs manipulating $N-1$ qubits of any fixed state from this
collection. In other words, there is a set of operators $U_{x}(m)$, including identity
operator, for which
\begin{equation}\label{45}
  m=N-1
\end{equation}
It is in agreement with (\ref{42}). It results in equation (\ref{44}) takes the form of
(\ref{41})
\begin{equation}\label{46}
\Phi_{b_{1}b_{2}\dots b_{N}}(N)=(\bbbone \otimes U_{b_{1}b_{2}\dots b_{N}}(N-1))\Omega
\end{equation}
where the string of bits $b_{1}b_{2}\dots b_{N}$ is binary notation of $x$, $x=0,\dots
2^{N}-1$.

Generally the question of existence and uniqueness of operators $U_{b_{1}b_{2}\dots
 b_{N}}(N-1)$ seems to be rather hard problem   and we shall discuss
 simple examples.  Let all operators be factorized  and have the form of product of the one
 particle operators
\begin{equation}\label{47}
U_{b_{1}b_{2}\dots b_{N}}(N-1)=U_{b_{1}b_{2}}(1)\otimes U_{b_{3}}(1)\dots U_{b_{N}}(1)
\end{equation}
Assume each of the transformations $U_{b_{1}b_{2}}(1), U_{b_{3}}(1)\dots$ can be
represented by the Pauli operators. If $N=2$ one finds
$U_{b_{1}b_{2}}(1)=\sigma_{x}^{b_{2}}\sigma_{z}^{b_{1}}$, and in accordance with
(\ref{44}) and (\ref{46})
\begin{equation}\label{48}
(\bbbone\otimes\sigma_{x}^{b_{2}}\sigma_{z}^{b_{1}})(|00\rangle+|11\rangle)= |0
b_{2}\rangle+(-1)^{b_{1}}|1 \bar{b}_{2}\rangle
\end{equation}
where $b_{1}, b_{2}=0,1$.  When $N>2$, the choice $U_{b_{k}}(1)=\sigma^{b_{k}}_{x}$
for $k=3,\dots N$ is suitable
\begin{eqnarray}\label{49}
(\bbbone\otimes\sigma_{x}^{b_{2}}\sigma_{z}^{b_{1}}\otimes\sigma_{x}^{b_{3}}\otimes\dots
\sigma_{x}^{b_{N}})\Omega&& \nonumber\\
=\frac{1}{\sqrt{2}}(|0 b_{2}\dots b_{N}\rangle +(-1)^{b_{1}}|1\rangle\otimes
|\bar{b}_{2} \dots\bar{b}_{N}\rangle)&&
\end{eqnarray}

The obtained equations (\ref{46}) and (\ref{49}) tell that the complete set of the $N$
qubit entangled states of the GHZ class can be associated with a set of the $N-1$
qubit operators, that generate all these states from one of them. In other words the
mapping given by (\ref{40}) can be justified. Indeed, the choice of operators may be
not unique. For example, if $N=3$, there is a case for which it is possible to
manipulate one qubit instead of two qubits
\begin{eqnarray}\label{491}
 (\bbbone \otimes\sigma_{z}\otimes\sigma_{x})(|001\rangle-|110\rangle) &&
 \nonumber
\\
 =(\bbbone\otimes\bbbone\otimes i\sigma_{y})(|001\rangle-|110\rangle) &&
\end{eqnarray}

The representation given by (\ref{46}) is not true for any states to be separable, it
needs entanglement not less then two particles. A state of $N$'s independent qubits
can be write in the form $\Phi_{b_{1}b_{2}\dots b_{N}}(N)=|b_{1}\dots b_{N}\rangle$.
When bits take their value 0 and 1, the obtained set is complete, but it is important,
that it can be generated from one of them by manipulating all  qubits. Then instead of
(\ref{46}), one finds $\Phi_{x}(N)=U_{x}(N)\Omega$. If two qubits are maximally
entangled and  others are independent, then such state has the form
$\Phi_{b_{1}b_{2}}(2)\otimes|b_{3}\dots b_{N}\rangle$, where $\Phi_{b_{1}b_{2}}(2)$ is
one of the Bell states. Entanglement allows to obtained complete set manipulating
$N-1$ qubits of an initial state, say, $\Omega'=\Phi_{00}\otimes|0\dots 0\rangle$. It
is important, that $\Omega'$ does not belong to the GHZ class by contrast $\Omega$,
given by (\ref{43}). From the physical point of view it is clear, that both states
$\Omega'$ and $\Omega$ can't be transformed from one to another by local operations.
For example, if $N=3$ one finds transformation
\begin{equation}\label{492}
(\bbbone\otimes C_{23})\Phi_{00}(2)\otimes|0\rangle=|GHZ\rangle
\end{equation}
where $\Phi_{00}(2)=(|00\rangle+|11\rangle)/\sqrt{2}$. Here the CNOT operation
$C_{23}$ involves two qubits simultaneously, that is an interaction between two
systems, that results in entanglement. When the GHZ is prepared, as initial state
$\Omega$, the complete set can be obtained in accordance with (\ref{46}), but
operators takes the nonlocal form $U_{b_{1}b_{2}b_{3}}=(\bbbone\otimes
U_{b_{1}b_{2}}(1)\otimes U_{b_{3}}(1))(\bbbone\otimes C_{23})$. This example indicate
the fact, that a complete set of the $N$ qubit entangled states can be generated
performing the non local operations on $N-1$ particles.

\subsection{Capacity of the channel}

Using (\ref{49}), one finds a dense coding scheme, that allows sending a $N$-bit
message by manipulating $N-1$ bits. To discuss capacity of the channel due from
entanglement it needs to replace $\Omega\to \alpha |0\rangle^{\otimes N}+\beta
|1\rangle^{\otimes N}$, where $|\alpha|^{2}+|\beta|^{2}=1$. Now the channel is not
assumed to be maximally entangled and its measure of entanglement, given by the
reduced von Neumann entropy, has the form
\begin{equation}\label{51}
E=-|\alpha|^{2}\log |\alpha|^{2}-|\beta|^{2}\log |\beta|^{2}
\end{equation}
The Holevo function reads $C(\{p_{x}\},\rho)=S(\rho /2^{N})$, where all probabilities
are equal and $p_{x}=1/2^{N}$. For the considered channel
\begin{eqnarray}\label{52}
\rho= \sum_{x}(\bbbone\otimes U_{x}(N-1))|\Omega\rangle \langle \Omega|(\bbbone\otimes
U_{x}(N-1))^{\dagger}&& \nonumber
\\
= \sum_{b_{1}\dots b_{N}=0,1}|\Phi'_{b_{1}b_{2}\dots b_{N}}(N)\rangle\langle
\Phi'_{b_{1}b_{2}\dots b_{N}}(N)|&&
\end{eqnarray}
Let operators $U_{x}(N-1)$ be factorized and have the form (\ref{49}), then
\begin{eqnarray}\label{53}
\Phi'_{b_{1}b_{2}\dots b_{N}}(N)=\alpha |0\rangle\otimes|b_{2}\dots
b_{N}\rangle&&
\\
\nonumber
 +(-1)^{b_{1}}\beta |1\rangle\otimes | \bar{b}_{2} \dots\bar{b}_{N}\rangle
 &&
\end{eqnarray}
All these states are generated from $\Omega$  by the local unitary transformations,
then their degree of entanglement is  $E$, given by (51).

If $\alpha=\beta=1/\sqrt{2}$, then $E=1$ and the channel is maximally entangled. In
the same time it implies the important fact, that the set of states becomes complete
in accordance with (\ref{aa}) and all these states can be well distinguishable by
measuring. For a maximally entangled channel the equation (\ref{52})is the condition of
completeness and density matrix $\rho$ takes the form $\rho=\rho(1)^{\otimes N}$,
where the single particle density matrix is $\rho(1)=\bbbone/2$.

When the channel can be not maximally entangled, one finds
\begin{equation}\label{54}
\rho=\rho'(1)\otimes\rho(1)^{\otimes (N-1)}
\end{equation}
where $\rho'(1)=|\alpha|^{2}|0\rangle\langle 0|+|\beta|^{2}|1\rangle\langle 1|$ is the
one particle density operator. Before calculating the classical capacity of the
channel, given by the Holevo function, note that it can be normalized per transmitted
bit. For the considered protocol there are $N-1$ bits, that Alice transmits  to Bob.
Then using (\ref{54}), capacity of the channel has the form
\begin{equation}\label{55}
 c=\frac{C(\{p_{x}\},\rho)}{N-1}=1+\frac{E}{N-1}
\end{equation}
It takes maximum $c_{max}=N/N-1$, when $E=1$. It means that entanglement results in
increasing of the classical capacity of the $N$ particle channel by $N/N-1$ times.
Indeed, this result is clear without calculating. If a channel permits sending of
$N$-bits of classical information manipulating $N-1$ qubits, then profit is $N/N-1$,
that is enhancement of the channel capacity per transmitted bits.

\subsection{Sufficient tight and other schemes}

The main resources, given by (\ref{11}), are sufficient also for teleportation of the
entangled states, that have the form $\zeta=\alpha|0\rangle^{\otimes (N-1)}
+\beta|1\rangle ^{\otimes (N-1)}$. The task can be accomplished by the $N$ particle
channel $\Omega$ and the $N/N-1$ bits of classical information per transmitted
particle, due from a $N$ particle measurement. The measurement involves  all particles
to be teleported and one particle from $\Omega$. It can be described by observable of
the form (\ref{24}), where $\pi^{\pm}\to(\pi^{\pm})^{\otimes(N-2)}$ \cite{VA}.

The presented teleportation and the dense coding schemes are based on the mentioned
resources to be sufficient and minimal for these tasks. This set of schemes we shall
name \emph{sufficient tight} schemes by contrast the other ones, that can be obtained
if some additional resources are permitted.

Let introduce the $k$ bit operators $En(k)$ and $Den(k)$ to be transformations of
entanglement and disentanglement. Their existence is the open question generally, but
in a particular case one finds $CNOT$ and the Hadamard gates to be useful. We assume,
that these operators $En(k)$ and $Den(k)$ transform any state of $k$ independent qubits
into entangled state and vice versa. Some modifications of schemes arise when these
operations are permitted.

It is well known, that operator $Ent(2)$, say of the form $Ent(2) =C_{12}$, plays the
key role in the one-bit teleportation, when an unknown qubit is entangled with ancilla
\cite{1bitel}. It results in one bit of classical information is needed for sending
the qubit. Indeed, the one bit protocol can be directly  generalized for teleportation
of two entangled qubits, for which two bits of information are required.

For dense coding schemes all modifications reduce to preparing of the channel state
$\Omega$ and revising of observable just as the way the considered GHZ channel.
Suppose, there is a collection of $N$ qubits, in which the $k$ particles are
independent and the remainder  $N-k$ qubits are entangled. Let only one qubit from
entanglement be in the receiver hand.  For preparing $\Omega$  it needs entanglement
of all particles, that can be achieved with the use of operator $En(k+1)$. It looks as
all operators $U_{x}(N-1)$ from a \emph{sufficient tight} scheme are replaced as
follows $U_{x}\to U_{x}\otimes Ent(k+1) $. Revising of observable or measurement is
another independent step. Assume, the $N$ bit message is already encoded by entangled
states $\Phi_{x}(N)$. Then before measuring, these states can disentangled by operator
$Den(n)$, that produces $n$ independent qubits, where $n\leq N$. The measurement
becomes more simple because  observable can be described by a set of states in which
not all qubits, or maybe all of them are independent. The cost of modification is
$U_{x}\to Den(n)\otimes U_{x}$. As result, the main revising of the dense coding
\emph{sufficient tight} schemes is
\begin{equation}\label{50}
  U_{x}\to Den(n)\otimes U_{x}\otimes Ent(k+1)
\end{equation}
In the case of the GHZ channel (\ref{50}) takes the form $U_{x}\to  H_{B}C_{BC}\otimes
U_{x}(2)\otimes C_{BC}$.

Both the entanglement and the disentanglement operators are useful for modification of
the \emph{sufficient tight} teleportation schemes for which there are some ways how to
prepare the channel state $\Omega$. As the $N$ - particle channel allows transmitting
perfectly only the $N-1$ particle entangled state of the form
$\zeta(N-1)=\alpha|0\rangle^{\otimes (N-1)} +\beta|1\rangle ^{\otimes (N-1)}$ then one
of the general idea of modification is disentanglement of the state to be teleported:
$\zeta(N-1)\to (\alpha|0\rangle+\beta|1\rangle)\otimes|N-2\rangle$, where $N-2$
particles in state $|N-2\rangle$ can be entangled with two ancilla qubits, say in the
EPR state, for preparing the quantum channel $\Omega$. Combining with disentanglement
operations it results in a collection of schemes, which based on one EPR pair and the
Bell state measurement as one of the initial resource, however the measurement will
involve not all particles to be teleported.

We illustrate the generalization of the one-bit teleportation protocol, considering
for simplicity  how  to transmit two entangled qubits. The task can be accomplished, if
an unknown state $\zeta=(\alpha|00\rangle+\beta|11\rangle)_{12}$ is entangled with an
EPR pair of the form  $\Omega=(|00\rangle+|11\rangle)_{AB}/\sqrt{2}$ as follows
$C_{A2}|\zeta\rangle_{12}\otimes|\Omega\rangle_{AB}$. Then the joint measurement of
the qubit 1 and 2 in basis $\pi_{1}^{\pm}\otimes |b\rangle_{2}$, $b=0,1$ projects the
remainder qubits $A$ and $B$ onto the state to be equal to the unknown state up to
unitary transformations. Note, here the non Bell state measurement allows teleporting
two entangled qubits by the 2 bits of classical information, however it is not the
LOCC protocol.

This work was supported in part by Delzell Foundation and Russian RFBR.

\end{document}